\documentclass[twocolumns,letterpaper]{aa}
\usepackage{graphicx}
\usepackage{txfonts}
\begin{document}
   \title{Line bisectors and radial velocity jitter from SARG spectra 
         \thanks{Based on observations made with the Italian Telescopio
                 Nazionale Galileo (TNG) operated on the island of La Palma
                 by the Centro Galileo Galilei of INAF (Istituto Nazionale
                 di Astrofisica) at the Spanish Observatorio del Roque del los
                 Muchachos of the Instituto de Astrof\'{i}sica de Canarias.}
         }
         
   \author{A. F. Mart\'{\i}nez Fiorenzano \inst{1}\fnmsep\inst{2}, 
           R. G. Gratton \inst{2},
           S. Desidera \inst{2},
           R. Cosentino \inst{3}\fnmsep\inst{4},
           \and M. Endl \inst{5}
          }

    \offprints{A. F. Mart\'{\i}nez Fiorenzano, 
    \email fiorenzano@pd.astro.it
    } 

    \institute{Dipartimento di Astronomia – Universit\`a di Padova,
               Vicolo dell'Osservatorio 2, I-35122, Padova, Italy
               \and
               INAF - Osservatorio Astronomico di Padova,
               Vicolo dell'Osservatorio 5, I-35122, Padova, Italy
               \and
               INAF - Osservatorio Astrofisico di Catania,
               Via S. Sofia 78, Catania, Italy
               \and
               INAF - Centro Galileo Galilei, Calle Alvarez de Abreu 70,
               38700 Santa Cruz de La Palma (TF), Spain
               \and
               McDonald Observatory, The University of Texas at Austin, 
               Austin, TX 78712, USA
               }

   \date{Received ; accepted }

   \abstract{
   
    We present an analysis of spectral line bisector variations for a few stars 
    observed in the SARG high precision radial velocity planet survey, and discuss their relationship
    with differential radial velocities. The spectra we consider are the same used for determining 
    radial velocities. The iodine cell lines employed in the measurement of radial velocities were removed 
    before bisector analysis. The line bisectors were then computed from average absorption profiles 
    obtained by cross correlation of the stellar spectra with a mask made from suitable lines of 
    a solar catalog. Bisector velocity spans were then determined: errors in these quantities 
    compare well with theoretical expectations based on resolution, S/N and line shape. The plot 
    of bisector velocity span against radial velocity was studied to search for correlations between 
    line asymmetries and radial velocity variations. A correlation was seen for HD 166435 due to 
    stellar activity, and for HD 8071B due to spectral contamination by the companion. No correlation 
    was seen for 51 Peg and $\rho$ CrB, stars hosting planets. We conclude that this technique may be 
    useful to separate radial velocity variations due to barycenter motion from spurious signals in 
    spectra acquired with the iodine cell.

   }

   \authorrunning {A. F. Mart\'{\i}nez Fiorenzano et al.}

   \maketitle


\section{Introduction}

  The study of activity jitter is mandatory in the search for exoplanets using the
  radial velocity (RV) technique, because it represents an important source of noise and a
  proper analysis is necessary to discard false alarms (e.g., HD 166435: Queloz et al. \cite{queloz}; 
  HD 219542B: Desidera et al. \cite{desidera03}, \cite{desidera04a}). The differential 
  RV variations induced by stellar activity are due to changes in the profile of spectral 
  lines caused by the presence of spots and/or the alteration of the granulation pattern 
  in active regions (Saar \& Donahue \cite{saar97}, Hatzes \cite{hatzes02}, Saar \cite{saar03} 
  and K\"{u}rster et al. \cite{kurster03}).
  The activity jitter of a star may be predicted by means of statistical 
  relations from its chromospheric emission, rotational velocity or amplitude of 
  photometric variations (Saar et al. \cite{saar98} and Paulson et al. \cite{paulson04}).
  Simultaneous determination of RV, chromospheric emission and/or photometry
  is even more powerful in disentangling the origin of the observed RV 
  variations Keplerian vs. stellar activity. However these techniques cannot be 
  considered as a direct measurement of the alterations of the spectral line profiles 
  that are the origin of the spurious RV variations.\\ 
  This type of study can be carried out by considering variations of line bisectors, that may 
  be thought of as direct measure of activity jitter (Queloz et al. \cite{queloz}) through the 
  evidence of variations of the asymmetries which appear in spectral lines. Line bisectors 
  are employed in the analysis of asymmetries (Gray \cite{gray82} and \cite{gray88}) due to 
  stellar atmospheric phenomena like granulation (Dravins et al. \cite{dravins}), turbulence or 
  pulsation that sometimes are responsible for the RV variations observed in stars.\\
  The reason for studying simultaneously the variations of the RV and of the (average) line bisector 
  of a star is to determine if any variation of the measured RVs are caused by a center of mass 
  motion due to celestial bodies orbiting a star (see Queloz et al. \cite{queloz} and 
  Desidera et al. \cite{desidera03}).\\
  The analysis of spectral line asymmetries also has an important application for the 
  follow-up of transit surveys. An eclipsing binary in a hierarchical triple system and a transiting 
  planetary companion might produce similar photometric signature; but very different line bisector 
  variations, allowing one to disentangle the characteristics of the system (Torres et al. \cite{torres04} 
  and \cite{torres05}).
  In addition, asymmetries of spectral lines may arise due to contamination of the spectrum by a 
  nearby star. This point is of particular relevance for the targets we are studying in the SARG 
  survey (Desidera et al. \cite{desidera04b}), because by design all of them are visual binaries. 
  In this case a companion near the line of sight may contaminate the spectral features of the star being 
  observed. Finally, spurious line profile asymmetries may be due to instrumental causes, e.g., 
  non symmetric illumination of the slit.\\ 
  Line bisector variations may be studied quite easily in spectra acquired using fibers 
  (see e.g., Queloz et al. \cite{queloz}). The fibers provide a constant, roughly symmetric illumination
  of the slit. Furthermore, spectra are generally acquired with simultaneous wavelength calibration lamps,
  rather than imprinting absorption cell features on the stellar spectral lines. No attempt 
  has been made to our knowledge to study line bisectors on spectra obtained through an iodine cell. 
  One disadvantage is the necessity to remove the iodine lines from the stellar spectra; but on the other 
  hand, the iodine lines allow a fine wavelength calibration and the possibility of monitoring the 
  instrumental profile.\\  
  In this paper we present such an attempt; line bisectors and radial velocities are determined for SARG 
  stellar spectra, to study possible trends between spectral line shapes (line bisectors) and RVs.\\
  This paper describes the procedure followed to handle the spectra, to remove iodine lines,
  the construction of a mask made from selected lines of a solar catalogue,
  the cross correlation function (CCF) computed between mask and stellar spectra,
  the determination of the line bisector and the calculation of the bisector 
  velocity span. We present results obtained for five stars: we found a clear correlation 
  between the line bisector velocity span and RV for HD 166435, likely due to stellar activity, 
  consistent with previous published work by Queloz et al. (\cite{queloz}). 
  A similar correlation was found for HD 8071B due to contamination by light from the companion. 
  We did not find any correlation for stars known to host planets like 51 Peg and $\rho$ CrB, 
  and the false alarm (inconclusive) case of HD 219542B.\\
  In the next section we describe some aspects of the observations; Section 3 explains the
  procedure employed in the analysis with results presented in Section 4. The last 
  two sections are devoted to a discussion of the error analysis and conclusions.\\ 

   \begin{figure}
   \centering
   \includegraphics[angle=90,width=8.9cm]{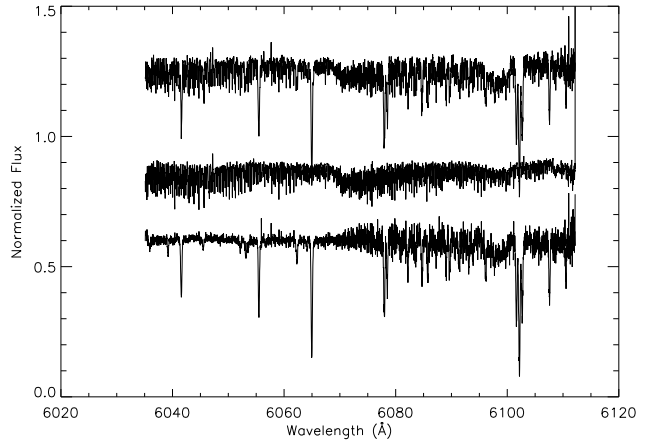}
      \caption{Portion of a spectrum from HD 166435.
              The top spectrum corresponds to the program star with iodine lines, in the middle 
              is the spectrum of the B-star with iodine lines and on the bottom is the program star 
              spectrum after the division by the B-star spectrum. The division was computed 
              for the whole order: the poor result is due to errors in the 
              wavelength calibration along one order. In this case the Iodine lines are better 
              removed in the left half than in the right half of the order. 
              }
         \label{fig1}
   \end{figure}
%

\section{Observations and data reduction}

   The data discussed in the present paper are part of the RV survey 
   aimed to find planets around stars in wide binaries (Desidera et al. \cite{desidera04b}),
   ongoing at TNG using the high resolution spectrograph SARG (Gratton et al. \cite{gratton01}). 
   The spectra have a resolution of $R \sim 150000$, covering the spectral range 
   4580\,\AA\, - 7900\,\AA\, in 55 echelle orders, with S/N values in the range 
   70 - 400 and were taken from September 2000 until August 2004. Slit width was set at 0.27\arcsec, much smaller
   than the typical seeing Full Width Half Maximum (FWHM). Furthermore, an autoguider system, viewing the slit 
   by means of a detector with its wavelength response peaked at the wavelength of the iodine cell lines, was 
   employed, keeping the instrumental profile stable and fairly independent of illumination effects. Guiding was 
   generally done using the image of the binary companion on the slit viewer.\\   
   Data reduction was performed in a standard way using IRAF\footnote[1]{IRAF is distributed
   by the National Optical Observatory, which is operated by the Association of Universities
   for research in Astronomy, Inc., under contract with the National Science Fundation.}. 
   High precision RVs were measured on these spectra with the AUSTRAL code (Endl et al. \cite{endl}) 
   as described in Desidera et al. (\cite{desidera03}).\\   
   The iodine cell technique includes the acquisition of spectra from a featureless source 
   (a fast rotating B-star or the flat field lamp) taken with the iodine cell inserted in the optical 
   path (Butler et al. \cite{butler96}). This kind of spectra is necessary only for 
   the deconvolution of stellar templates (taken without the cell), but they were acquired also 
   in most of the observing nights of our survey to monitor instrument performances.
   These featureless source spectra were used to remove the Iodine lines from the science spectra,
   as explained in the next section.\\

\section{Data analysis}  
     
   \begin{figure}
   \centering
   \includegraphics[width=8.9cm]{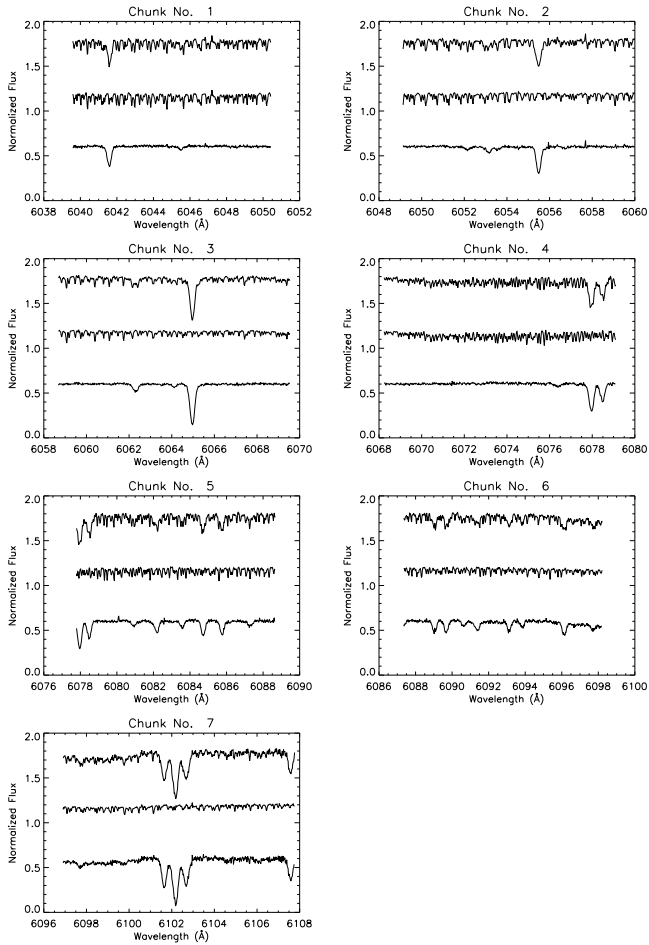} 
      \caption{An order of a spectrum of HD 166435 (wavelength range $6039 \AA - 6107\AA$)
               shown in Figure ~\ref{fig1}. The order has been divided into 7 chunks, 
               as described in the text.
               The spectra are shifted arbitrarily in flux to show: on top the stellar 
               spectra with iodine lines, in the middle the B-star spectra (adjusted to 
               the wavelength offset) with the iodine lines and on the bottom the clean stellar 
               spectra after dividing the program star spectrum by the B-star spectrum. 
              }
         \label{fig2}
   \end{figure}
%
\subsection{Handling of the spectra}

   The iodine lines, superimposed on the stellar spectrum for wavelength
   calibration and RV determination, were eliminated because the 
   line bisector is intended to study the asymmetries of the stellar
   spectral lines alone. For this task, spectra of fast rotating 
   ($V\,sin\,i\geq 200\,\rm{km\,s^{-1}}$) B-stars were employed. The analysis was made only 
   for the wavelength range where the iodine lines appear in the spectra 
   (5036\,\AA\, - 6108\,\AA\,) along 21 orders. We used only this spectral region because there 
   the wavelength calibration is more accurate, because of the iodine lines themselves.\\
   To handle the spectra, each order was divided in 7 pieces of 500 spectral points each, 
   corresponding to a wavelength width of $\sim 10 \,\AA$, overlapped by 60 points ($\sim 1\,\AA$) 
   to eventually recover any absorption lying at the edges of the chunks. This procedure of 
   cutting the spectra in pieces was clearly advantageous, while the division of complete 
   orders displayed not optimal results, related to errors in the wavelength calibration of 
   the spectra (see Figure ~\ref{fig1}).   
   A cross correlation computed between the spectral chunks of the B-star and the program star 
   determined the offset in wavelength between the iodine lines common to both spectra. 
   The B-star spectrum was adjusted using a Hermite spline interpolation (INTEP, see Hill \cite{hill}) 
   to the wavelength scale defined by the program star spectrum after application of the appropriate 
   offset and finally divided to the program star spectrum (see Figure ~\ref{fig2}; details will appear 
   elsewhere: Mart\'{\i}nez Fiorenzano \cite{martinezfiorenzano}). 
   The success of this procedure depends on the stability of the instrumental profile over time, since 
   the B-star spectra were usually acquired at the beginning or the end of each night.\\
   In a few cases, the cross correlation procedure used to determine the wavelength offset 
   of chunks did not provide a reasonable value; in these cases the division of the star flux by 
   the B-star flux added noise rather than removing the iodine lines. These noisy chunks were rejected from 
   further analysis.\\
   Spikes due to cosmic rays or hot pixels were removed by replacing the spectral values within them 
   with the averaged flux of adjacent spectral points. This procedure was only applied to those cases where 
   spikes occurred far from the line centers. In the very rare cases where the spikes occurred near the 
   centers of the lines used in our analysis, possibly affecting the line bisector analysis described below, 
   the relevant chunks were simply removed from further analysis, from all spectra of the same star.\\

   \begin{figure}
   \centering      
   \includegraphics[height=18cm, width=8.9cm]{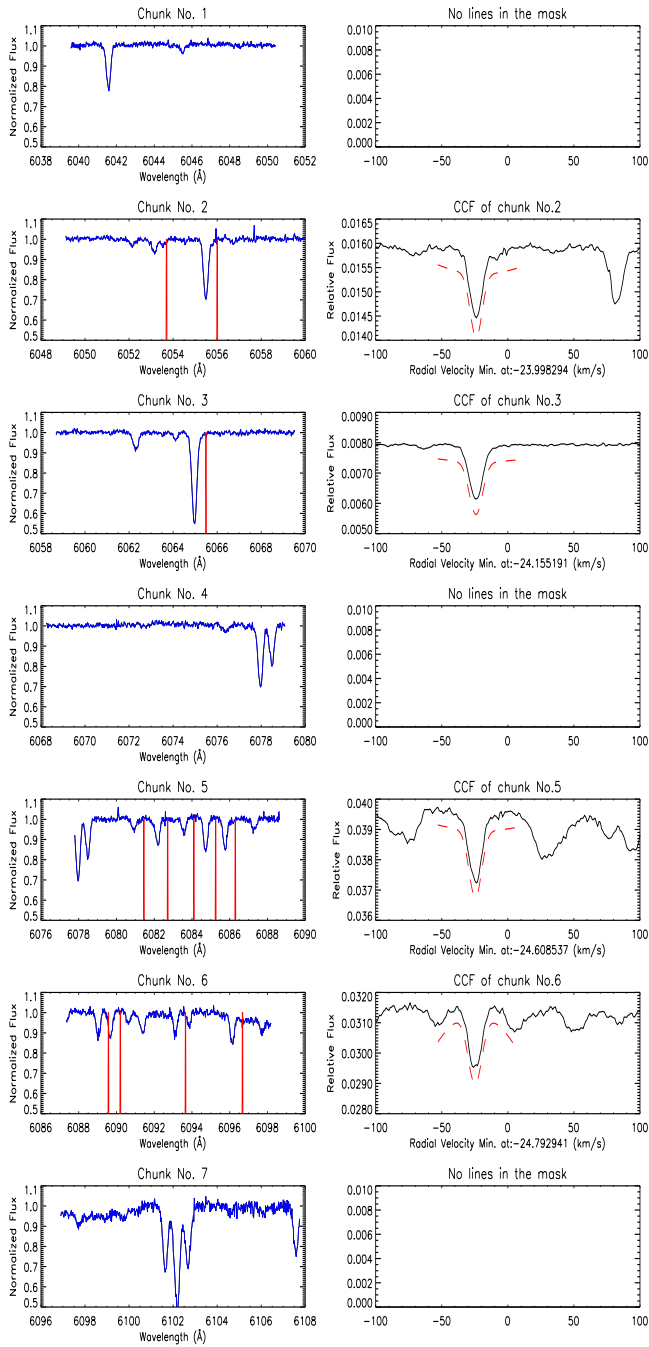}
      \caption{An order of a spectrum of HD 166435. The left column shows individual
              chunks of the spectrum (blue lines), after removal of the iodine cell
              lines (see Figure ~\ref{fig1}). Overimposed (as red lines) is the mask used for the
              determination of the CCF (only lines classified as $very\,\,good$ are shown
              here). It is made of a sum of $\delta-$functions centered on the rest
              wavelengths of the selected lines. Note the wavelength shifts between the
              stellar lines and the mask peaks due to the non-zero radial velocity of
              the star. The CCF for individual chunks computed by cross-correlating the
              spectra with the masks given on the left panels are shown in the right
              panels. The dashed line, shifted slightly below the profiles for clarity, 
              represent the Gaussian fit computed to determine the local minimum of the 
              CCF for each chunk. There are no CCF profiles for chunks 1, 4 and 7 due to 
              the lack of suitable lines for the mask in these wavelength ranges.      
              }
         \label{fig3}
   \end{figure}
\subsection{The cross correlation function (CCF)} 

   \subsubsection{The solar catalogue and line selection for the mask}

   The list of spectral lines by Moore et al. (\cite{moore}) was used to prepare the spectral mask 
   needed for the computation of the CCF. A preliminary line list was obtained by selecting those 
   lines that do not have possible contaminants (wavelength separation $\gtrsim 0.1 \,\AA$), and have 
   reduced width between 3 and 30 F. F (Fraunhofer) is defined as the dimensionless quantity 
   $\Delta \lambda/ \lambda \times 10^{6}$, where $\Delta\lambda$\ is the equivalent width 
   (see Moore et al. \cite{moore}). The range of reduced width chosen here corresponds to a range of 
   central line depths from 0.14 to 0.75 in continuum units (in the solar spectrum). The CCFs we derived
   thus represent average profile for lines of intermediate strength.\\
   Once a preliminary list of lines was determined for the wavelength range of interest 
   (5036\,\AA\,\,to 6108\,\AA), a further selection was made by inspecting the ``The Sacramento 
   Peak Atlas of the Solar Flux Spectrum" (Beckers et al. \cite{beckers76}). Lines were 
   labeled according to their appearance: ``y" ($very\,\,good$): sharp, clear, with weak wings; 
   ``y:" ($good$): clear but near other lines altering their wings; 
   ``?" ($not\,\,very\,\,good$): with small blends or strong wings. Many other lines, 
   blended or too close to other spectral lines, were removed from the list at this step.\\
   The mask used to compute the CCF is a sum of $\delta$-function profiles corresponding 
   to 1 for the line wavelength, with a base of two (spectral) points, and 0 elsewhere.
   The extensive line list (including $very\,\,good,\,\,good$ and $not\,\,very\,\,good$ lines: 
   a total of about 300 lines) served to determine the RV; this gave the centroid value of the average 
   absorption profile obtained by the CCF. Once this point was established, which helps to properly 
   locate the profiles for the addition, a new computation of the CCF was performed only with the 
   $very\,\,good$ lines to obtain an improved average profile, used to calculate 
   reliable average line bisectors (see Figure ~\ref{fig3}).\\ 

   \subsubsection{The cross correlation and addition of profiles}
 
   A cross correlation between the mask and the stellar spectrum was computed
   for each chunk; the addition of all these cross correlations gave the average profile. 
   Due to the relatively low S/N of some spectra, use of the average of many lines is appropriate 
   for this study (the variations of line bisectors with time). On the other hand, 
   the actual line bisector depends on the line depth, so that 
   our ``average" line bisector does not rigorously correspond to the line bisector of a line with 
   similar depth of the CCF. Therefore, the use of our average profile would be misleading for some 
   scientific goals such as the study of convection in stellar atmospheres.\\
   Due to different illumination of the CCD, each chunk along an order has different flux 
   values, those close to the center of the orders being more luminous and yielding then results of higher S/N. 
   To account for this, before summing the individual cross correlations, each of them was multiplied by an 
   appropriate weight, proportional to the instrumental flux at the center of the chunk.\\
   Once adjusted to a common reference frame, which may be thought of as a centering procedure, and 
   multiplied by its weight, all CCF profiles were added to get the final profile for the spectrum.
   This was normalized to determine the finally adopted average line profiles using the reference 
   continuum determined in the IRAF reduction. This was obtained by interpolating a polynome throughout the 
   spectra with a suitable clipping rejection procedure. Note that this reference continuum may contain 
   significant errors, so that there may be points in the normalized CCFs that are well above unity. 
   However, we kept our procedures strictly uniform throughout the analysis of different spectra; we
   then expect that these errors in the location of the reference continuum mainly affect absolute 
   profiles and bisector estimates, but only marginally their spectrum-to-spectrum variations, which
   are of interest for the present discussion.\\

\subsection{The line bisector calculation}
   
   The bisector of an absorption line is the middle point of the horizontal segment 
   connecting points on the left and right sides of the profile with the same flux level. 
   The line bisector is obtained by combining bisector points ranging from the core toward the 
   wings of the line.\\ 
   For the bisector determination, it is necessary to adjust the ordinate axis of the profile 
   to a convenient scale and to determine the values in velocity belonging to the left and right 
   points for a given flux value. This is accomplished by interpolating (using INTEP, see Hill \cite{hill}) 
   the absorption profile to the desired scale (details will appear in 
   Mart\'{\i}nez Fiorenzano \cite{martinezfiorenzano}).\\

   \begin{figure}
   \centering
   \includegraphics[angle=0,width=9cm]{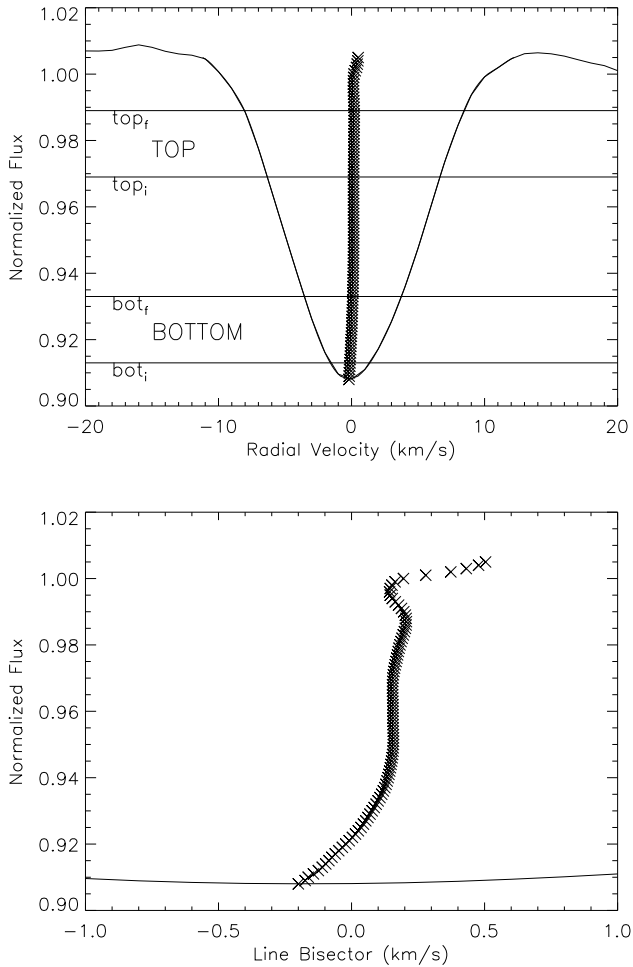}
      \caption{Spectrum of HD 166435. In the top panel we show the normalized cross correlation 
               profile, the line bisector, the top and bottom zones (both with $\Delta F = 0.02$; 
               $\Delta F = top_{f} - top_{i} = bot_{f} - bot_{i})$. 
               In the bottom panel we show a zoom of the profile with the RV scale 
               increased to better display the asymmetries of the line bisector.
              }
         \label{fig4}
   \end{figure}
%
\subsection{The bisector velocity span}
   
   To quantify the asymmetry of the spectral lines and look for correlation with RV it is
   useful to introduce the bisector velocity span (Toner \& Gray \cite{toner}). This is 
   determined by considering a top zone near the wings and a bottom zone close to the core 
   of the lines, which represent interesting regions to study the velocity given by the bisector 
   (see Figure ~\ref{fig4}).
   The difference of the average values of velocities in the top and bottom zones, 
   $V_{T}$ and $V_{B}$ respectively, determine the bisector velocity span (BVS for short).\\ 
   The location of the top and bottom zones, as well as their width $\Delta F$, were determined 
   as percentages along the absorption profiles. They were defined using as templates the spectra 
   of HD 166435; the same zones were then employed in the study of the other objects 
   to be consistent in the analysis procedure.\\
   The Spearman nonparametric rank correlation coefficient was computed to establish the 
   significance of the linear correlation between BVS and RV for each star.\\

\subsection{Error determination}
    
   The expression given by Gray (\cite{gray83}) for the bisector error due to 
   photometric error was employed to establish the error for the bisector velocity 
   span. The photon noise error, determined for the specific case of SARG using 
   Gray's equation, yields the expression:
   
\begin{equation}
      \delta V = \left( \frac{S}{N} \right)^{-1} 
                 \left( 2nF \frac{\Delta F}{x} \frac{dF}{dV} \right)^{-1/2} \,,
\end{equation}  

\noindent    
   where $S/N$ is the signal-to-noise of the spectrum, including the contribution of the B-star spectrum
   and it is given by: $({(N/S)_{star}^2 + (N/S)_{B-star}^2})^{-1/2}$, with values between 65 - 300. 
   The B-star spectra employed in the analysis were acquired on the same observing nights of the 
   target spectra and have typical S/N values $\gtrsim 250$. In eq. (1) $n$ is the number of lines in 
   the mask for the CCF; $F$ is the central flux of the zone of analysis; $\Delta F$ is the interval 
   of flux determining the zone; $x$ is the scale of the spectrum based on the linear dispersion of the 
   instrument ($1.04\,\rm{km\,s^{-1}\,pix^{-1}}$ for SARG) and $dF/dV$ gives the slope of the profile in 
   the zone being analyzed (see Figure ~\ref{fig4}).\\
   Error bars for the BVS were then determined by adding quadratically the errors for 
   the top and bottom zones.\\

\subsection{Instrument Profile asymmetries}

\begin{figure}
   \centering
   \includegraphics[angle=0,width=9cm]{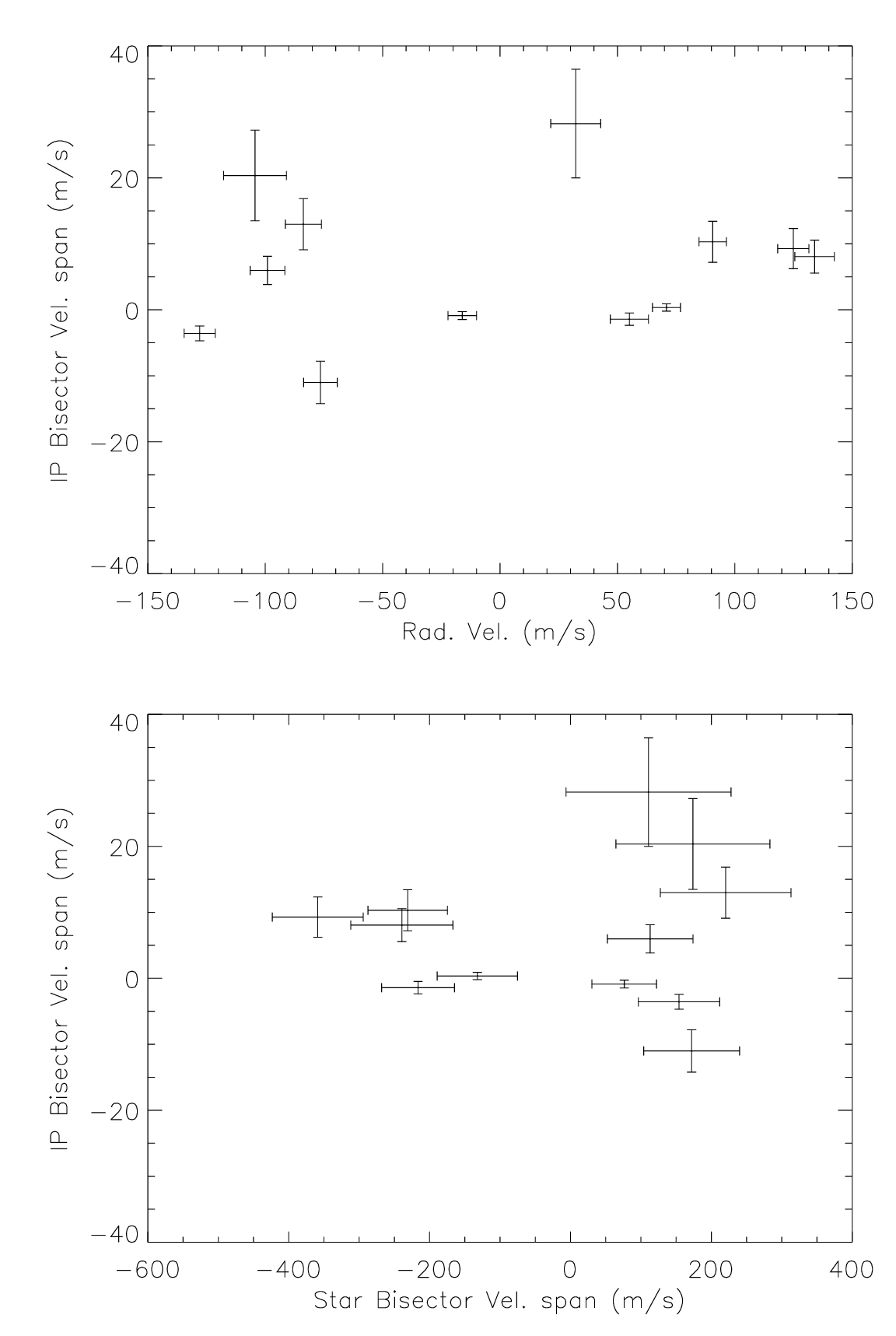}
      \caption{ Instrumental profile (IP) analysis. Top Panel: plot of IP BVS vs. RVs of HD 166435.
              Bottom panel: plot of IP BVS vs. BVS of HD 166435. No evidence of trends
              were found.  
              }
         \label{fig5}
   \end{figure}

   Line bisector variations may arise from variations of the stellar spectra as well as from variations 
   of the instrumental profile (IP). We studied the IP determined by AUSTRAL as part of the RV determination 
   and computed line bisectors for it as for stellar spectra.\\ 
   Generally the line bisector variations of IP are significant, but they are typically much lower than 
   the errors in the line bisectors determined from the stellar spectra. Corrections due to asymmetries of 
   line profiles require complex convolution, which should take into account the real stellar profile. 
   Actually, since the IP is much narrower than the intrinsic stellar line profile 
   (the case of SARG spectra), corrections due to IPs are much smaller than the IP asymmetries 
   themselves; therefore we did not apply corrections to the stellar BVS.\\
   In Figure ~\ref{fig5} we show the run of the BVS of the IP with RV and the IP BVS with the stellar 
   BVS in the analysis of HD 166435. No clear trend could be discerned.\\

%

   \begin{figure}
   \centering
   \includegraphics[angle=0,width=9cm]{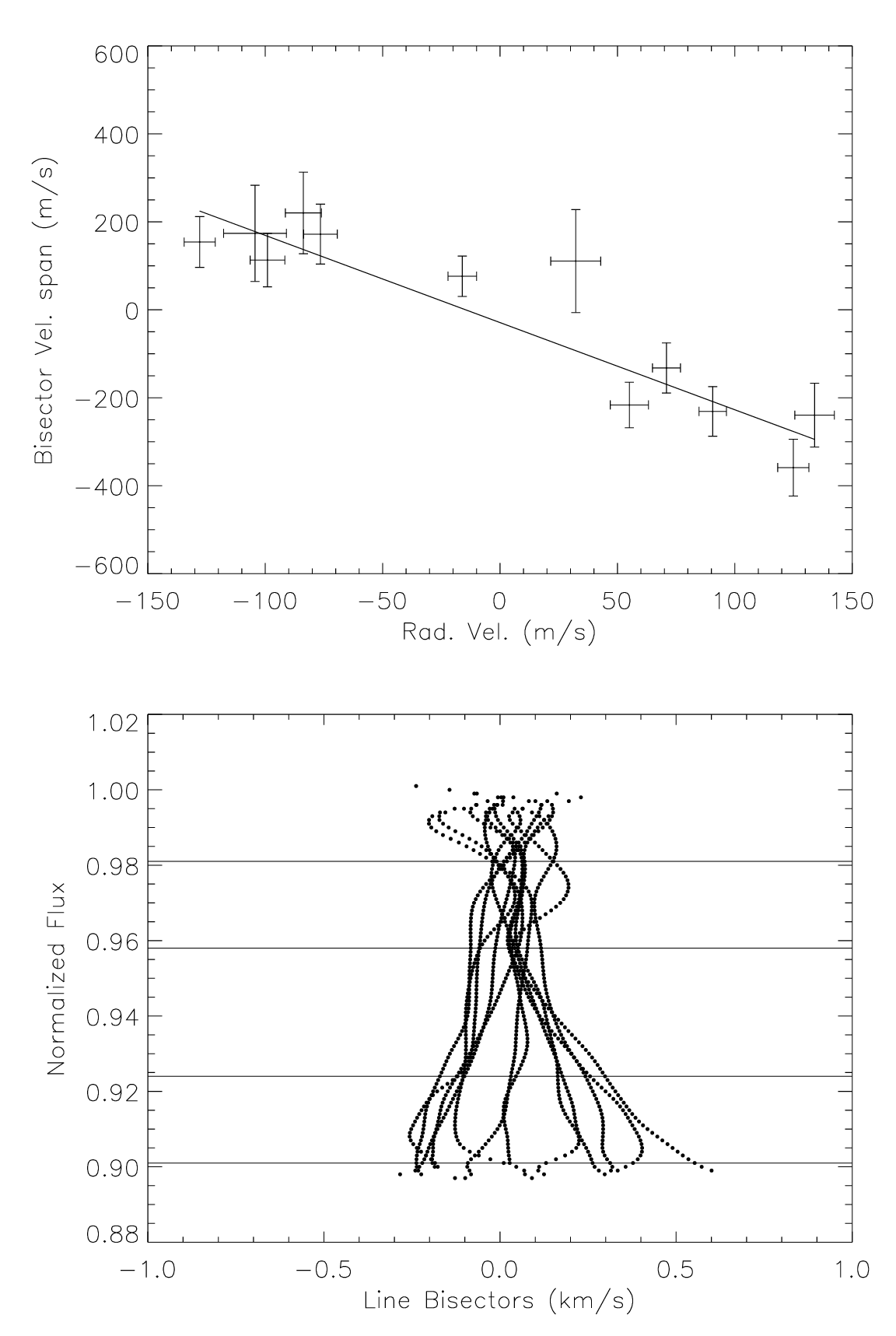}
      \caption{Upper panel: plot of BVS vs. RV for HD 166435 and 
               the best fit to a straight line. Lower panel: line bisectors for individual
               spectra adjusted to their corresponding RV. The horizontal lines enclose 
               the top and bottom zones considered for the fitting analysis.
               }
         \label{fig6}
   \end{figure}
%
   \begin{table}
         \caption[]{Bisector velocity span from spectra of HD 166435. 
                    53 lines were employed in the mask for the CCF.}
         \label{tab1}
        $$ 
         \begin{array}{crrc}
            \hline
            \noalign{\smallskip}
            \rm{JD - 2450000} & (V_{T}-V_{B})   & V_{r}\,\,\,\,\,\,\,\,\, & \rm{S/N}\\
                         & \rm{m\,s^{-1}}\,\,\,\,\,\, & \rm{m\,s^{-1}}\,\,\,\, &    \\
            \noalign{\smallskip}
            \hline
            \noalign{\smallskip}
                       
 2775.65  & -231.1\pm  56.4  &   90.6\pm   5.8  & 164 \\
 2776.68  &  154.1\pm  57.8  & -127.9\pm   6.6  & 156 \\
 2809.62  & -216.4\pm  51.8  &   55.1\pm   8.2  & 178 \\
 2818.62  &  173.8\pm 109.3  & -104.4\pm  13.4  &  86 \\
 2860.40  &  220.3\pm  92.8  &  -83.8\pm   7.7  &  98 \\
 2861.42  &  110.8\pm 117.1  &   32.3\pm  10.6  &  81 \\
 2862.42  & -358.8\pm  64.5  &  124.9\pm   6.6  & 140 \\
 2891.37  &  172.0\pm  68.1  &  -76.5\pm   7.2  & 138 \\
 2892.35  & -239.4\pm  72.5  &  134.0\pm   8.4  & 126 \\
 3129.74  &  113.1\pm  60.8  &  -99.0\pm   7.4  & 149 \\
 3130.73  &   76.3\pm  45.9  &  -16.1\pm   6.1  & 201 \\
 3131.74  & -132.2\pm  57.0  &   70.9\pm   6.0  & 164 \\
                       
            \noalign{\smallskip}
            \hline
         \end{array}
      $$
       
   \end{table}

\section{Results}
     
\subsection{HD 166435}

   The dwarf star HD 166435, of spectral type G0 and visual magnitude $m_{V} = 6.8$, shows
   evidence of RV variations, photometric variability and magnetic activity. Furthermore, 
   previous analysis of the variation of the line bisectors revealed a correlation between RV and 
   line bisector orientation (Queloz et al. \cite{queloz}).\\
   The large amplitude of the activity-induced variations and their stability make this
   star an ideal target to test new procedures like those presented in this paper.\\
   Twelve spectra of HD 166435 were acquired between May 2003 and May 2004.
   The line bisectors, derived from the spectra, were used to study the correlation 
   between the BVS and the RVs. The plot of BVS against RV ($(V_{T} - V_{B})\,\,vs.\,\,V_{r}$) 
   was fitted to a straight line.\\
   In principle, the location and extension of top and bottom zones of the line profiles used in the 
   derivation of the BVS are arbitrary. In order to optimize detection of variations of the BVS 
   correlated with RV variations, we maximized the $m/\sigma$ ratio (where $m$ is the slope of the 
   straight line fitting of BVS against RV, and $\sigma$ its uncertainty) over different choices of
   the location and amplitude of the line profile regions used to derive the BVS. The $m/\sigma$ ratio 
   is an indicator of the significance of the linear term in the straight line fitting.\\
   The top and bottom zones were determined according to the relative absorption percentages in 
   which the highest significance was found: top centered at $25\%$ of the maximum absorption, 
   bottom at $87\%$; in both cases the width $\Delta F$ was of $25\%$. 
   For consistency, these percentage values were then considered for the analysis of all the stars.\\
   The upper panel of Figure ~\ref{fig6} displays the plot of BVS against RV; superimposed is the best 
   fitting straight line, with:
\begin{equation}
      (V_{T} - V_{B}) = (-1.98\pm 0.21)\times V_{r} + (-29.03\pm 18.45) \,.
\end{equation}
    
\noindent  
   The rank correlation coefficient is $-0.89$ and the significance is of $99.99\%$. 
   The lower panel of the same figure shows the line bisectors computed for all spectra of HD 166435, 
   corrected for their RV and plotted on a common reference frame. Values of bisector velocity span and 
   RV for individual spectra are listed in Table ~\ref{tab1}. The trend of the BVS obtained in this work
   is similar to that of Queloz et al. (\cite{queloz}); the larger value we found for the slope 
   can be attributed to the higher resolution of the SARG spectra.      
   \begin{figure}
   \centering
   \includegraphics[angle=0,width=9cm]{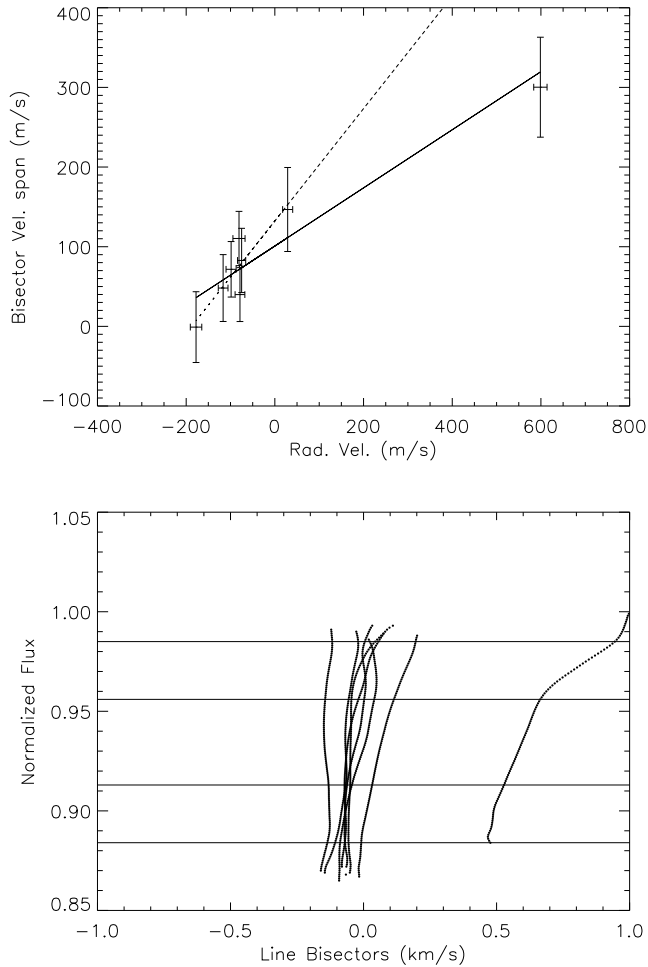}
      \caption{Upper panel: plot of BVS vs. RV for HD 8071B and 
               the best fit to a straight line. 
               The dotted line corresponds to the
               best fit discarding the spectrum with higher RV. 
               Lower panel: line bisectors for individual
               spectra adjusted to their corresponding RV. The horizontal lines enclose 
               the top and bottom zones considered in the fitting analysis.
               }
         \label{fig7}
   \end{figure}
%
   \begin{table}
         \caption[]{Bisector velocity span from spectra of HD 8071B. 
                    87 lines were employed in the mask for the CCF.}
         \label{tab2}
     $$ 
         \begin{array}{crrc}
            \hline
            \noalign{\smallskip}
            \rm{JD - 2450000} & (V_{T}-V_{B})   & V_{r}\,\,\,\,\,\,\,\,\, & \rm{S/N}\\
                         & \rm{m\,s^{-1}}\,\,\,\,\,\, & \rm{m\,s^{-1}}\,\,\,\, &    \\
            \noalign{\smallskip}
            \hline
            \noalign{\smallskip}

 1797.64  &  110.3\pm  34.0  &  -80.9\pm  13.7  & 137 \\
 1854.54  &  146.7\pm  52.8  &   28.8\pm  11.3  &  91 \\
 2145.71  &   71.6\pm  34.9  &  -98.7\pm  11.4  & 140 \\
 2297.38  &   -1.0\pm  44.4  & -177.7\pm  12.9  & 110 \\
 2892.61  &   48.1\pm  42.0  & -116.6\pm  10.9  & 112 \\
 2982.48  &  300.3\pm  62.8  &  598.7\pm  15.0  &  92 \\
 3216.73  &   39.9\pm  33.8  &  -78.7\pm  11.0  & 141 \\
 3246.68  &   82.7\pm  40.3  &  -74.9\pm   9.2  & 121 \\
                       
            \noalign{\smallskip}
            \hline
         \end{array}
     $$
   \end{table}

   \begin{figure}
   \centering
   \includegraphics[width=9cm]{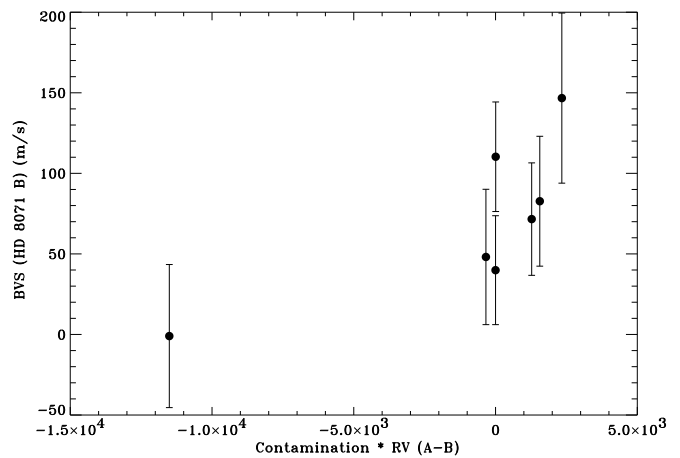}
      \caption{ BVS of HD 8071B vs. the ``effective contamination'' of RV, 
                defined as the product of the light contamination from the companion HD 8071A 
                and the RV difference between the components, variable with time 
                because HD 8071A is itself a spectroscopic binary. The highly discrepant
                spectrum from JD: 2452982.48 was excluded.
                Abscissa axis has 
                dimensions of $(percentage ~of ~contamination)\times(m/s)$.
              }               
       \label{fig12}
   \end{figure}
%

\subsection{HD 8071B}

   The star HD 8071B of spectral type G0V and visual magnitude $m_{V} = 7.6$ is a member 
   of a visual binary system. The primary HD 8071A is brighter by 0.3 mag and has a projected 
   separation of 2.1\arcsec, the smallest in the sample of the SARG survey.
   Pairs with such small separation are typically 
   not observed in cases of bad seeing; furthermore the slit was oriented perpendicularly
   to the separation of the components. Nevertheless, some contamination of the spectra
   is still possible, especially for the secondary component.
   Our velocity measures show that HD 8071A is itself a single lined spectroscopic binary with 
   semiamplitude of $\sim 7\,\rm{km\,s^{-1}}$. The RV difference between the visual components 
   is then variable in the range (-4, 10) $\rm{km\,s^{-1}}$.\\ 
   Eight spectra of HD 8071B were acquired between September 2000 and August 2004. 
   The correlation between BVS and RV found for all spectra has rank correlation 
   coefficient of $0.81$ and significance of $98.5\%$. The analysis of the IP did not show any critical trend 
   acting on the RV computation, leading us to consider contamination as the cause of the observed 
   correlation. The spectrum with RV $\sim 600\,\rm{m\,s^{-1}}$ and very different line profile is likely 
   heavily contaminated by the companion, because of the occurrence of technical problems related to 
   telescope tracking that night. Our study of the line bisector allows us to clearly identify the problematic 
   spectrum and to exclude it from the analysis of the radial velocity curve.\\   
   Nevertheless a correlation still persists even without considering the highly discrepant spectrum, with a rank 
   correlation coefficient $0.71$ and significance of $92.9\%$. This is likely due to some residual contamination 
   (of a few percent at most) by the companion, compatible with the small separation of the pair and the actual 
   observing conditions. Figure ~\ref{fig7} shows the two cases and the eight line bisectors computed for all 
   spectra of HD 8071B, corrected for their RV and plotted on a common reference frame. Values of bisector 
   velocity span and RV for individual spectra appear in Table ~\ref{tab2}.\\
   To confirm our hypothesis of contamination as the source of observed BVS-RV correlation, 
   we performed a simple modeling of the expected contamination, excluding the highly discrepant
   spectrum from JD: 2452982.48. We first determined the light contamination expected on the spectrum of HD 8071B on the 
   basis of the seeing conditions (given by the FWHM of the spectrum measured along a direction perpendicular 
   to the dispersion).
   RV and BVS of HD 8071B do not show significant correlation with such contamination.
   We then considered an ``effective contamination'' as the product of the light contamination and the 
   variable RV difference between HD 8071 A and B. The Spearman rank correlation coefficient of RV and
   BVS vs. such ``effective contamination'' is 0.85 with a significance of more than 96\%, supporting our
   hypothesis of light contamination as the origin of the RV and BVS variability of HD 8071B.
   The relation between RV perturbation and BVS is not linear likely because the RV perturbation due to 
   contamination is expected to be a non-linear function of the position of the contamination across the 
   line profile (See Figure ~\ref{fig12}).\\

   \begin{figure}
   \centering
   \includegraphics[width=9cm]{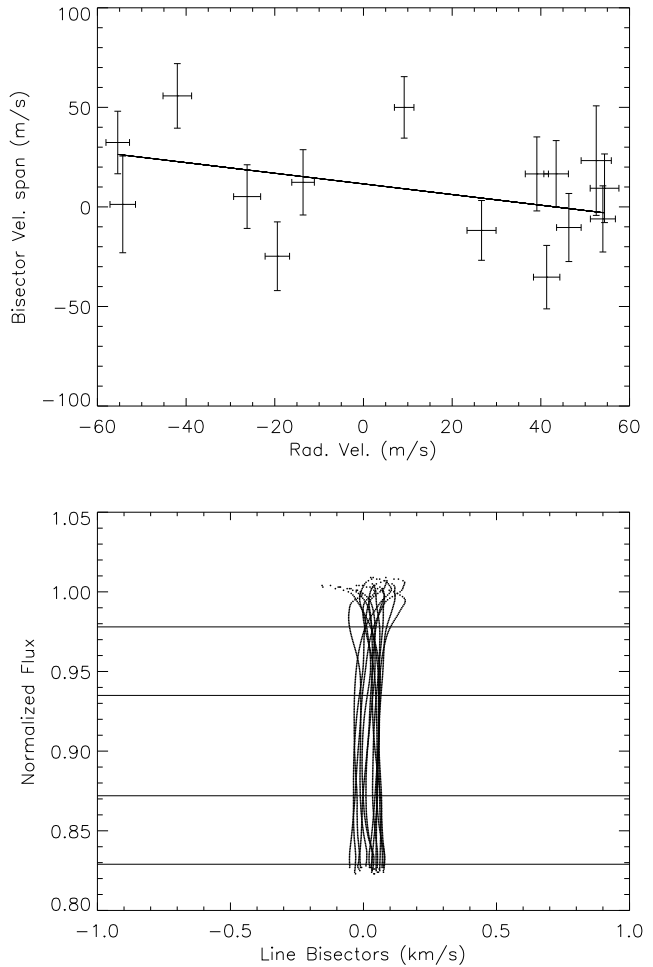}
      \caption{Upper panel: plot of BVS vs. RV for 51 Peg and 
               the best fit to a straight line. Lower panel: line bisectors for individual
               spectra adjusted to their corresponding RV. The horizontal lines enclose 
               the top and bottom zones considered for the fitting analysis. 
               74 lines were employed in the mask for the CCF.
               }
               
       \label{fig8}
   \end{figure}
%

\subsection{51 Peg}

   The star 51 Peg (HD 217014) of spectral type G2.5IVa and visual magnitude $m_{V}=$ 5.5  
   is the first discovered solar-type object to host a planet, with $M_{2}$ sin$i$ $= 0.47\,M_{J}$, 
   $a = 0.052$ AU and period of $4.23$ days (Mayor \& Queloz \cite{mayor}).\\ 
   The star lies in a zone of the Hertzsprung-Russell diagram of stable (very low variability) objects 
   (Eyer \& Grenon \cite{eyer}). Indeed, according to Henry et al. (\cite{henry00}), no measurable 
   change in mean magnitude (over 5 years) was seen and the Ca II record displayed a signal essentially constant 
   despite some season-to-season jitter and a general indication of a low activity level.\\ 
   Fifteen spectra of 51 Peg were acquired between June 2001 and November 2003. 
   There is no significant correlation between BVS and RV (rank correlation coefficient of $-0.28$ and 
   significance of $56\%$), confirming the results by Hatzes et al. (\cite{hatzes98}) and 
   Povich et al. (\cite{povich01}). 
   The line bisector shape (see Figure ~\ref{fig8}) seems to be constant.\\ 

   \begin{figure}
   \centering
   \includegraphics[angle=0,width=9cm]{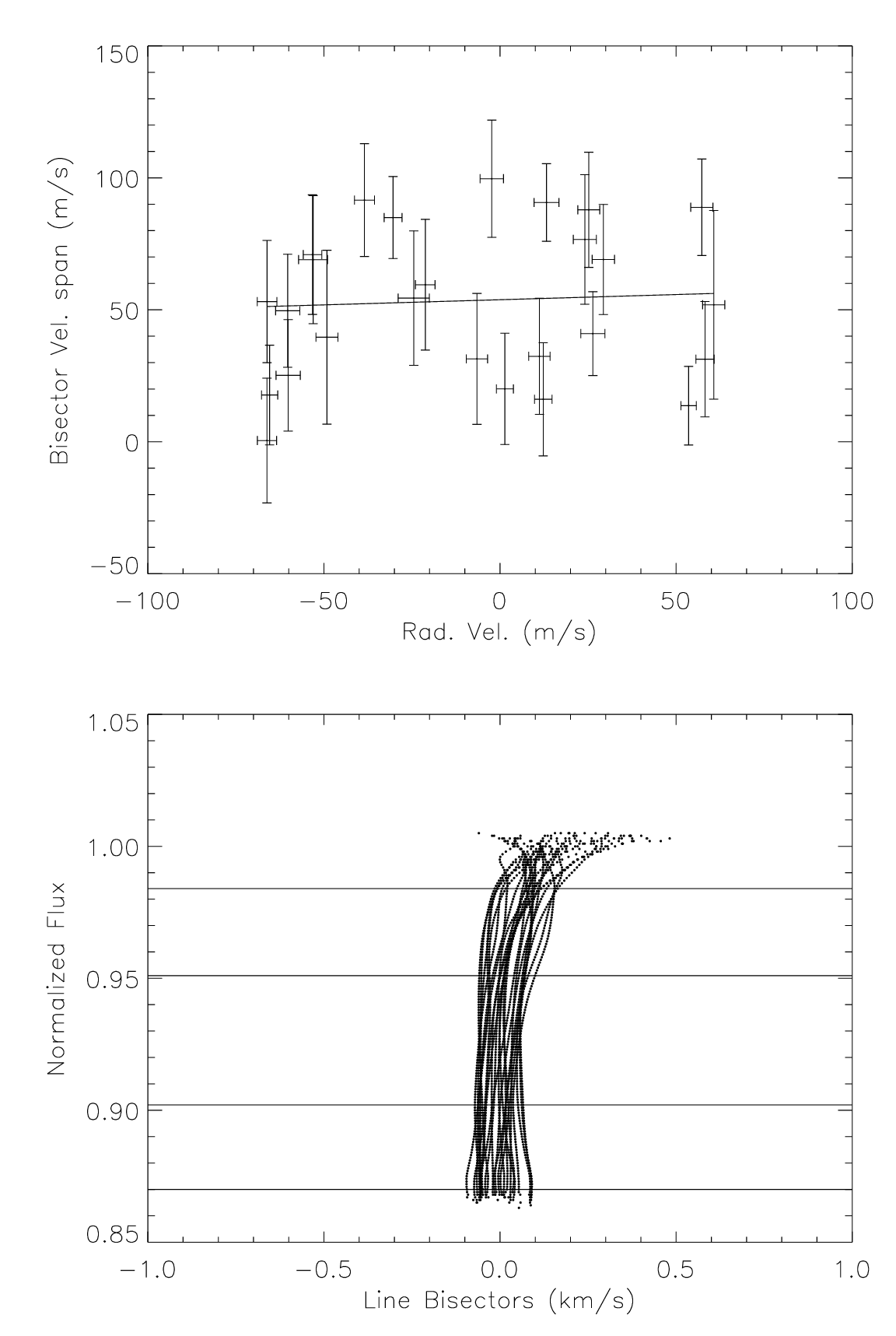}
         \caption{Upper panel: plot of BVS vs. RV for $\rho$ CrB and 
               the best fit to a straight line. Lower panel: line bisectors for individual
               spectra adjusted to their corresponding RV. The horizontal lines enclose 
               the top and bottom zones considered for the fitting analysis.
               85 lines were employed in the mask for the CCF.
               }
         \label{fig9}
   \end{figure}
%

\subsection{$\rho$ CrB}

   The star $\rho$ CrB (HD 143761) of spectral type G0Va and visual magnitude $m_{V}=$ 5.4 is known 
   to host a planet with $M_{2}$ sin$i$ $= 1.04\,M_{J}$, $a = 0.22$ AU and a period of $39.95$ days 
   (Noyes et al. \cite{noyes97}).\\
   Twenty six spectra of $\rho$ CrB were acquired between April 2001 and March 2004.
   In the plot of BVS against RV the dispersion of the data points shows no correlation (as in Povich 
   et al. \cite{povich01}) with a rank correlation coefficient of $0.15$ and significance of $52\%$.\\
   The line bisectors and its behaviour, similar to those for 51 Peg, are shown in Figure ~\ref{fig9}. 
   The typical ``C" shape of line bisectors is more evident for $\rho$ CrB in agreement with the warmer 
   temperature of the star.\\

\subsection{HD 219542B}

\begin{figure}
   \centering
   \includegraphics[angle=0,width=9cm]{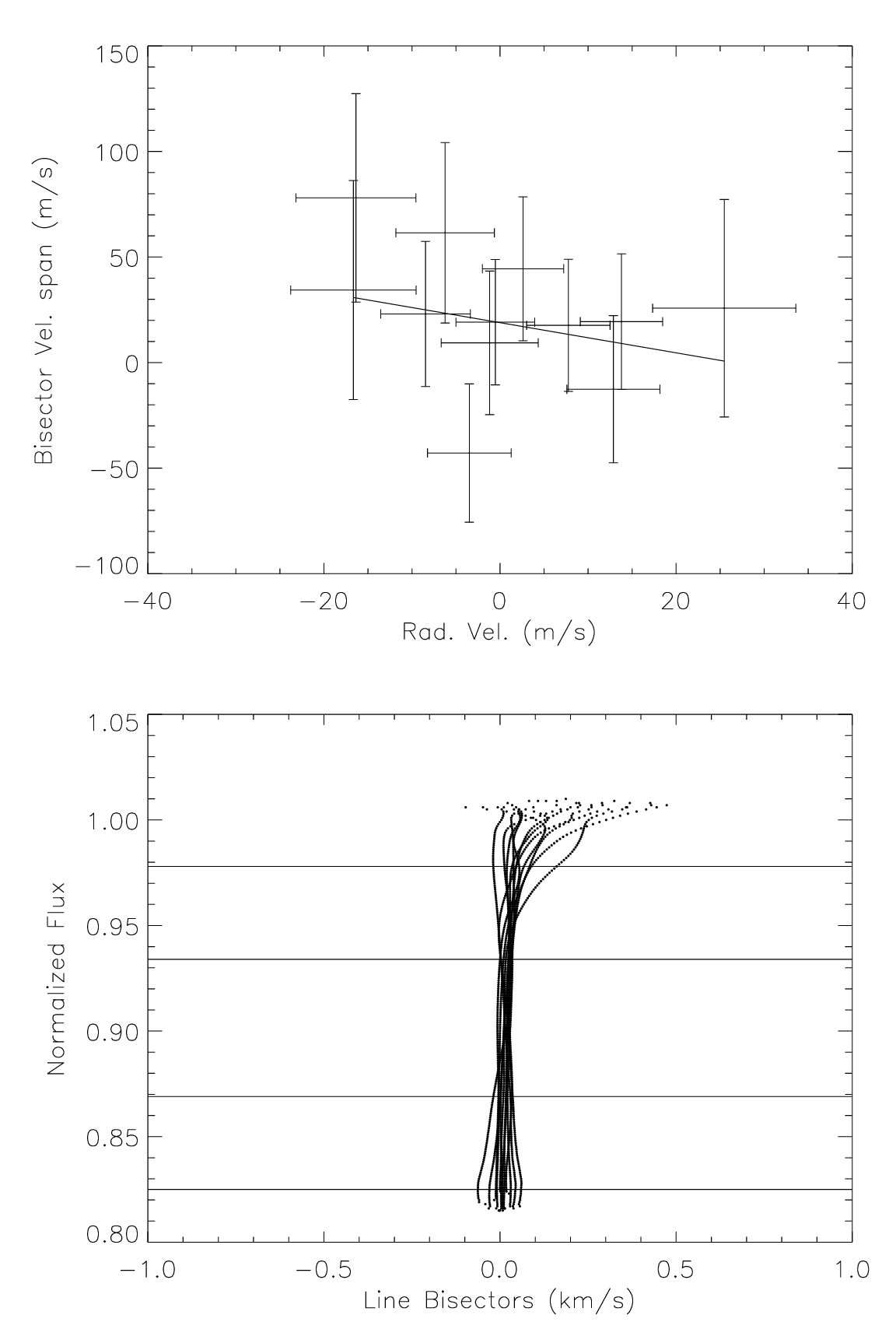}
         \caption{Upper panel: plot of BVS vs. RV for HB 219542B and 
               the best fit to a straight line. Lower panel: line bisectors for individual
               spectra adjusted to their corresponding RV. The horizontal lines enclose 
               the top and bottom zones considered for the fitting analysis.
               86 lines were employed in the mask for the CCF.
               }
         \label{fig10}
   \end{figure}

   The star HD 219542B is member of a wide binary system, with spectral type G7V 
   and visual magnitude $m_{V}=8.6$. It was considered in Desidera et al. (\cite{desidera03}) as a
   candidate to host a planet, but ultimately discarded after further analysis; the small RV 
   variations are more likely related to a moderate stellar activity (Desidera et al. \cite{desidera04a}).\\
   For the present analysis of the line bisectors, only the data of the 2002 season were considered,   
   twelve spectra, because the RV scatter and chromospheric activity were greater in this season 
   (see Desidera et al. \cite{desidera04a}). Therefore it should be easier to find a correlation between
   BVS and RV from these data alone.\\ 
   The plot of BVS against RV and the line bisectors are shown in Figure ~\ref{fig10}. 
   No correlation appears in this case: the rank correlation coefficient is of $-0.37$ and the significance 
   of $76\%$. This lack of correlation is due to the small velocity amplitude 
   (approximately between -17 and 26 $\rm{m\,s^{-1}}$) and to the low S/N of the available spectra $\sim 100$.\\   

%
   \begin{figure}
   \centering
   \includegraphics[angle=90,width=7.9cm]{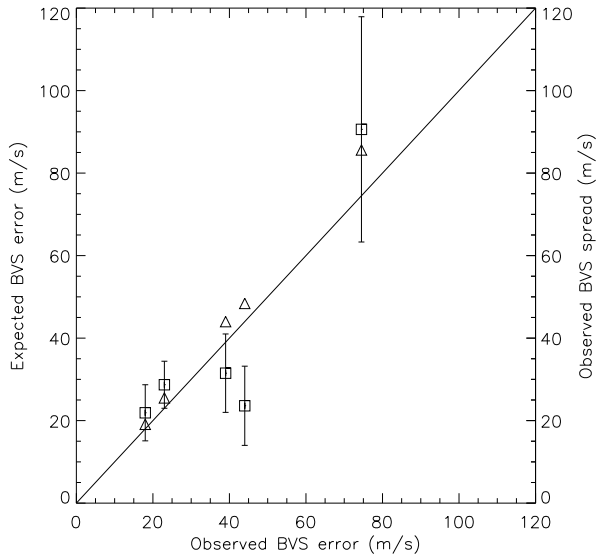}
      \caption{The expected against observed errors of BVS 
               (triangles: quadratic mean of error bars).
               Squares correspond to observed BVS spread against the measured BVS errors 
               with spread errors as error bars.                 
              }
         \label{fig11}
   \end{figure}
  

\section{Error analysis}

  An analytic study of errors was performed starting by considering the internal BVS errors. 
  The expected errors of BVS were computed for absorption 
  profiles obtained by convolution of Gaussian and rotational profiles. The former were determined 
  with a thermal broadening factor estimated by colors and temperatures of each star and the 
  latter with a $V\,sin\,i$ factor determined by the Fast Fourier Transform (FFT) analysis 
  of each star's absorption profile (Gray \cite{gray92}).\\ 
  The observed errors of BVS were the quadratic mean of the error bars of the single spectra.
  The observed BVS spread was the rms average of the individual BVS for the stars that do not
  show a significant BVS vs. RV correlation (51 Peg, $\rho$ CrB and HD 219542B), and the rms of residuals
  from the linear correlation for HD 166435 and HD 8071B.\\
  The values of errors estimated by these procedures are given in Table ~\ref{tab6}. The plot 
  in Figure ~\ref{fig11} compares the observed spread of data with expectations based on internal
  errors alone. These two set of values are fully consistent with each other, indicating that our 
  procedure is not affected by large systematic errors.\\
  
   \begin{table}
      \caption[]{BVS errors: computed by eq. (1) in convolved profiles (expected) 
                 and measured from CCF profiles (observed).
                }
         \label{tab6}
     $$ 
         \begin{array}{lccc}
            \hline
            \noalign{\smallskip}
            \rm{Star} &  \rm{BVS\,error} & \rm{BVS\,error}    & \rm{BVS\,spread} \\
                 & \rm{observed}\,\rm{(rms)}   & \rm{expected}  & \rm{observed} \\
                 &   \rm{m\,s^{-1}}  & \rm{m\,s^{-1}}     & \rm{m\,s^{-1}}  \\
            \noalign{\smallskip}            
            \hline
            \noalign{\smallskip}

 \rm{HD\,\,166435}  & 74.5 & 85.6 &  90.6^{a}\pm 27.3 \\
 \rm{HD\,\,8071B}   & 44.0 & 48.4 &  30.9^{a}\pm 11.7 \\
 \rm{51\,\,Peg}     & 18.0 & 19.1 &  25.3\pm  6.8 \\
 \rm{\rho\,\,CrB}   & 23.0 & 25.5 &  28.7\pm  5.7 \\
 \rm{HD\,\,219542B} & 39.0 & 44.0 &  31.5\pm  9.5 \\
                       
            \noalign{\smallskip}
            \hline
         \end{array}
     $$
\begin{list}{}{}
\item[$^{\mathrm{a}}$] These values correspond to the rms of residuals from the linear correlation.
\end{list}
   \end{table}
   

\section{Conclusions}

  We studied the variation of line bisectors in the same spectra acquired through the iodine cell, employed 
  also for high precision RV measurements. We found that such variation, as measured by the BVS, 
  shows spreads fully consistent with internal errors, as determined from photon statistics,
  spectral resolution and intrinsic line profiles. We correlated the variation of the bisector 
  span with high precision RVs for five stars.\\
  A significant correlation was established in two cases: an anticorrelation for HD 166435, 
  as found by Queloz et al. (\cite{queloz}). This is due to the stellar variability 
  and magnetic activity, which makes the core of the profiles change from positive 
  to negative values of RV. A positive correlation for HD 8071B was due to contamination of the 
  spectra by light from the companion star producing an asymmetry in the red wings of the profiles 
  with a consequent inclination of the line bisectors toward positive values of RV.\\
  For the stars known to host exoplanets, 51 Peg and $\rho$ CrB, no correlation was found, further 
  supporting the conclusion that in these two cases RV variations are due to Keplerian motion.\\
  The special case of HD 219542B shows that the bisector technique does not allow us to disentangle the
  activity origin of low amplitude RV variations ($\sim 20\,\rm{m\,s^{-1}}$) using spectra of S/N $\sim 100$ 
  even at the high resolution of SARG spectra ($R \sim 150000$).\\
  We conclude that spectra acquired using the iodine cell may be used to study variations of line bisectors. 
  In order to achieve the required accuracy, it is necessary to deal with high quality spectra, in particular 
  high S/N to reduce the error bars in the BVS, or to study spectra where the RV variations are of large amplitude.\\

\begin{acknowledgements}
       We thank the referee Dr. Steven H. Saar for his very detailed report that helped to improve 
       the paper. A. F. Mart\'{\i}nez Fiorenzano thank Mikhail Varnoff for support and advice.    
       This work was partially funded by COFIN 2004 ``From stars to planets: accretion, 
       disk evolution and planet formation" by Ministero Universit\`{a} e Ricerca 
       Scientifica Italy.\\ 
\end{acknowledgements}


\end{document}